# Structural limitations of learning in a crowd:

# Communication vulnerability and information diffusion in MOOCs.

Nabeel Gillani[1], Taha Yasseri[2], Rebecca Eynon*[2], Isis Hjorth[2]

[1] Department of Engineering Science University of Oxford Parks Road Oxford OX1 3PJ UK
nabeel@robots.ox.ac.uk

[2] Oxford Internet Institute University of Oxford 1 St. Giles Oxford OX1 3JS UK {taha.yasseri, rebecca.eynon, isis.hjorth}@oii.ox.ac.uk

**Abstract:**

Massive Open Online Courses (MOOCs) bring together a global crowd of thousands of learners for several weeks or months. In theory, the openness and scale of MOOCs can promote iterative dialogue that facilitates group cognition and knowledge construction. Using data from two successive instances of a popular business strategy MOOC, we filter observed communication patterns to arrive at the "significant" interaction networks between learners and use complex network analysis to explore the vulnerability and information diffusion potential of the discussion forums. We find that different discussion topics and pedagogical practices promote varying levels of 1) "significant" peer-to-peer engagement, 2) participant inclusiveness in dialogue, and ultimately, 3) modularity, which impacts information diffusion to prevent a truly "global" exchange of knowledge and learning. These results indicate the structural limitations of large-scale crowd-based learning and highlight the different ways that learners in MOOCs leverage, and learn within, social contexts. We conclude by exploring how these insights may inspire new developments in online education.

Over the last few years, millions of self-selected learners have enrolled in courses on large-scale learning platforms, typically co-participating with thousands of peers in courses of their choice[1]. Within the field of Computer Supported Collaborative Learning (CSCL), the structural properties of online group learning have been studied in detail[2,3]. However, these studies have been carried out in the contexts of more traditionally organized, smaller-scale online classrooms. The novelty of "learning at scale" has inspired a number of research studies that explore how participants interact with course content. Some studies have revealed that these courses are taken primarily by those already with college degrees[4]. Others have traced the behavior of MOOC participants through a course's lifespan by leveraging granular "clickstream" data[5,6]. Still others have started to cluster learners based on their patterns of engagement in order to predict dropouts[7].

Despite a growing body of research[8,9], many questions relating to the characteristics of group interactions and dialogue in these courses have largely been ignored. Contemporary socio-cultural learning theory emphasizes the role of group interaction for cognition, highlighting the need for understanding the degree to which MOOCs in practice allow for deep and meaningful learning through the facilitation of significant interactions and the spread of information between participants as they seek to acquire and generate new knowledge[10,11].

In this paper, we aim to expand this area of research by examining group level interactions in MOOC forums. Regardless of the type of MOOC, learners are typically offered the opportunity to participate and collaborate with one another in online discussion forums to enhance their educational experience. The forum participants have a diverse range of backgrounds and motivations for taking the courses. There are few rules, and are best defined as "non-formal" learning spaces[12], where learners are free to pick and choose how and if they interact. The overall governance structure is relatively weak, set primarily by the educators' questions / assignments for the forums, the technical design of the forums (e.g. the division of forums into specific topics) and the roles participants themselves take on during the course (typically 8 weeks).

While theoretical perspectives and emphases differ in studies of online learning, it is recognised that understanding the learning process in online forums requires consideration of interactions at the individual and group level[2,13,14]. The interactions at the group level within these forums can be viewed as a kind of scaffold through which learning can occur[15], and therefore, is of significant practical concern when considering the future design and development of courses.

A key challenge in addressing group-level interaction in MOOCs is of methodological character: we need to determine what constitutes a "significant" interaction in these learning environments of unprecedented scope. In large-scale MOOC forums, with socio-culturally diverse learners, this is a non-trivial problem. Many of the tens of thousands of interactions in the forum may have little relevance to enhancing the learning process; thus, the first question we address asks: *how can we determine the underlying social networks that depict significant interactions?*

Once the significant interactions have been established, we ask two questions core to understanding group interactions for learning in MOOCs: *What is the vulnerability of the communication networks in the forums, and how does information flow through the forums?* The large scale and open nature of the forums means that the group has a high set of resources available in the form of people with varied experiences and expertise. However, this kind of group membership, together with the non-formal and short-term nature of these courses, means that relatively weak inter-personal relationships are likely[15,16].

## Results

We analysed data from two successive instances of a business MOOC (FOBS-1 and FOBS-2), offered on the Coursera platform in Spring and Autumn of 2013. Nearly 90,000 students registered for FOBS-1 and over 77,000 for FOBS-2. The courses lasted for six weeks each, during which 4,500 FOBS-1 and 3,300 FOBS-2 learners contributed over 15,000 and 14,000 posts or comments in the discussion forums, respectively. More than 15,000 and 11,500 learners viewed at least one discussion thread in both instances, contributing to 181,911 and 139,858 total discussion thread views in both instances of the course.

The course's discussion forum was divided by course staff into multiple sub-forums before the start of the course. Some examples of sub-forums included a Cases sub-forum to facilitate weekly discussions about business cases and a Final Project sub-forum to facilitate questions and collaborations on the Final Strategic Analysis assignment (see the Methodology for more details). We analysed the dynamics of each sub-forum separately for a more granular understanding of learner behaviours. This approach was justified by low overlap between learners across sub-forums (less than 10% in all instances, and in most cases, below 3%). These numbers are revealing in their own right: most students participated infrequently in only a few sub-forums.

Both posting and viewing in each sub-forum tended to occur in bursts, with most activity happening at the beginning of the course, and in some cases, triggered by recurring or final course milestones (see Fig. S1 and S2 in Supplementary Information). The number of views and posts per discussion thread had a fat-tailed distribution (see Fig. S3 in Supplementary Information) and the threads of different sub-forums tended to have different "view lifespans"; the time period that 90% of views occurred within (Fig. S4 in Supplementary Information). Given our interest in exploring explicit forum participation, we focus the remainder of this work on posting behaviour. Table 1 provides basic summary statistics for the different sub-forums in the course (see the Methodology for an explanation of the sub-forums).

## Significant interaction networks

We modelled communication between learners as a social network, where nodes represent learners that posted explicitly in the discussion forums and an edge between two learners indicates that they co-participated in at least one discussion thread (see Supplementary Information for formal definition). However, not all links generated this way are equally important, and two learners' co-participation in a thread is not necessarily indicative of a meaningful social exchange. Drawing upon work from ecology and engineering[17], as well as recent work in mobile communication networks[18], we assumed that the observed communication network in each sub-forum was a noise-corrupted version of the "true" network – i.e., one that depicts meaningful communication between students. Our task, then, was to derive this true network by filtering out links between learners thought to have been generated by random encounters. Intuitively, the algorithm filtered out links that appeared to occur by chance - e.g., two users that co-participated a small number of times in a particular thread even though they were active participants in other parts of the sub-forum. Conversely, edges formed between students that participated in iterative, consistent dialogue were generally detected as significant. As Fig. 1 illustrates, filtering out insignificant edges impacted various network attributes, such as modularity[19]. See "deriving significant interaction networks" in the Methodology section for more details on the link filtration procedure.

Table 2 shows the proportion of edges pruned for the significant network derivation of each top-level sub-forum in FOBS-1 and FOBS-2. In both instances, the feedback forums had the greatest decrease in density, which was to be expected: most participation in these sub-forums was one-off, posting either specific technical questions/comments or a final "thank you" to the course staff at the end of the class. On the other hand, the Cases sub-forum lost the smallest proportion of edges at 44% and 39% in both instances, respectively. This is likely due to the type of discussion encouraged in this sub-forum: the weekly case discussions asked students to provide their thoughts and opinions on a series of open-ended questions. The sub-forums accommodated these questions and aimed to incite, and guide, group discussion. In many cases, students would read the analyses posted by their peers and comment with additional insights or critiques, leading to greater engagement and knowledge construction.

Interestingly, more than 2 out of 3 connections in the study groups sub-forum were considered "insignificant" in both instances of the course. Informal content analysis of this sub-forum revealed that most learners posted to introduce themselves early on in the course, particularly to their peers in similar geographic regions/time zones, then opting to move their discussions to other online platforms (e.g. Facebook). Largely, the Study Groups sub-forum served more as a meeting point for learners instead of an environment for sustained community-building and engagement.

## Communication vulnerability

The vulnerability of networks has been studied across disciplines[20]. For example, power systems engineers often ask which "critical set" of network components must be damaged in a functioning circuit in order to cut off the supply of electricity to the remaining nodes[21]. We asked an analogous question from an educational perspective: which "critical set" of learners is responsible for potential information flow in a communication network - and what would happen to online discussions if the learners comprising this set were removed? We defined vulnerability in the education context to be the proportion of nodes that must be disconnected from the network in order to rapidly degrade the relative size of the largest connected component to the total number of nodes. To our knowledge, the vulnerability of communication networks in educational settings has not been explored in previous research.

The vulnerability of MOOC discussion networks indicates how integrated and inclusive communication is. Discussion forums with fleeting participation tend to have a small proportion of very vocal participants comprise this set: removing these learners from the online discussions would rapidly eliminate the potential of discussion and information flow between the other participants. Conversely, forums that encourage repeated engagement and in-depth discussion among participants have a proportionally larger critical set, and discussion is distributed across a wide range of learners. By analysing vulnerability in different sub-forums, we sought to understand how group communication dynamics differed according to the topics being discussed.

To chart the vulnerability of each sub-forum, we executed the following algorithm: for each sub-forum's derived significant network, iteratively find the node with the highest betweenness centrality[30], disconnect it from the network, and compute the resultant proportion of nodes in the network's largest connected component to the total number of nodes. We used betweenness centrality as our removal metric as its definition, particularly for undirected graphs, indicated each node's potentiality as a conduit for the spread of information or ideas. We compared results from this removal strategy to one where nodes are removed at random, which has served as a baseline evaluative mechanism for similar analyses in other application domains[20].

Fig. 2 depicts the results for all of the sub-forums in FOBS-1 (with similar results for FOBS-2 illustrated by Fig. S5 in the Supplementary Information). Fig. 3 zooms into the Cases sub-forum from FOBS-1 to depict the degradation resulting from iteratively removing the node with the highest betweenness centrality to a random removal strategy. From Fig. 2, it is clear that different sub-forums had different vulnerability thresholds. For example, in both instances of the course, the Cases and Final Project sub-forum were the least vulnerable, as determined by the relatively higher proportion of nodes requiring removal before the relative size of the largest connect component was driven close to zero. This likely resulted from the high levels of iterative dialogue and knowledge construction characteristic of both sub-forums. Indeed, learners tended to use these forums to share ideas and insights as they related to the weekly case questions, their final strategic analyses, or questions about course outcomes and the peer review process[25]. These trends, which could be interpreted as conducive to promoting learners' participation in multiple discussions with many other learners, were largely absent in other sub-forums. For example, in the Study Groups sub-forum – one of the most vulnerable – the majority of individuals posted once to introduce themselves to a particular group of students and then proceeded to move their discussions to other platforms, or perhaps, cease engagement altogether. Indeed, it is interesting to note differences in sub-forum vulnerability across both FOBS-1 and FOBS-2. The proportion of nodes required to rapidly degrade both the Cases and Final Project sub-forums is higher in FOBS-2 than in FOBS-1, suggesting less communication vulnerability. This may have been the result of an additional evaluation criteria in FOBS-2, where 8% of students' final scores was computed as a function of the total number of "upvotes" they received on their posts or comments in the discussion forum.

The different vulnerability thresholds across sub-forums and course iterations suggest that the different topics being discussed, and perhaps, different incentives for participation promoted different levels of inclusiveness and engagement among learners. A movie of a simulated degradation for the FOBS-1 Final Project sub-forum is also included in the Supplementary Information.

## Information diffusion

From analyzing network vulnerability, it is clear that different sub-forums have different "critical sets" of participants that characterize the inclusiveness of discussions. Still, it is important to explore how information spreads in these networks, as doing so may ultimately reveal how forum participation promotes knowledge construction. Therefore, we ask: *how does information flow through the forums*?

To investigate this, we simulated an information diffusion model similar to the SI (Susceptible-Infected) model of contagion[22,26] which has been extensively used in previous work to model social contagion[24,27].

Although very simplistic, the SI model is very useful in analyzing the topological and temporal effects on networked communication systems. The SI model does not take into account effects such as decaying interest over time, the influence of peers, and more sophisticated mechanisms of social contagion, but it adequately determines the upper limit of the contagion rate based on the topology and the connectedness of the interaction network. It also enables us to compare different topologies and their efficiency in information spread within a quantitative framework. Please see the Supplementary Information for more details.

As a benchmark, we performed the same diffusion simulation on a randomized network, where each node maintained its degree but had a different set of neighbors than those observed in the significant network (i.e., a configuration model[23]). The purpose of shuffling neighbors in the randomized network is to present the diffusion potential of the corresponding sub-forum *without inherent modularity* – i.e., the benchmark provides a baseline of how well information would flow between participants in the forum if the observed community structures did not exist. The time evolution of the percentage of reached nodes both for the original and randomized networks is depicted in Fig. 4(a). It is evident from this figure that the spread is uniformly faster in the randomized networks, and throughout the process starting from a single infected node until the whole system is infected. To quantify the difference between the randomized and original networks, we computed the time it took for a simulated "information packet" to come into contact with half of the network's nodes (i.e., participants), $T_{\text{half}}$. It is possible to consider a different threshold for prevalence (e.g., some have used 20%[27]), but as evident from Fig. 4(a) the choice of this threshold does not change the overall patterns.

We defined an information diffusion efficiency $e$ of a sub-forum as $e = T_{\text{half (random)}} / T_{\text{half (original)}}$. Values of $e$ that were larger/smaller than 1 indicated that the structure of the discussion network correlated with enhanced/diminished information diffusion compared to the randomized benchmark. Fig. 4(b) illustrates the value of *e* for the different sub-forums of FOBS-1. Similar trends have been observed for FOBS-2 (depicted in Fig. S6 in the Supplementary Information).

Across all sub-forums, information spreads more slowly in the original networks compared to the randomized benchmarks. This is likely explained by the existence of local community structures in the derived significant networks. It has recently been shown that a dense cluster of nodes with a large number of connections could impede diffusion processes[24]. The Questions for Professors sub-forum was the most efficient at facilitating information spread; conversely, the highly modular Study Groups sub-forum depicted the least diffusion efficiency (the Methodology section explains the intended function of each of the depicted sub-forums).

To further support this finding, we calculated the modularity score[28] of the original and randomized networks of each sub-forum, using the Louvain community detection method[19] with resolution 1[29]. We then computed the normalized modularity score for each sub-forum by taking the ratio of the original network's modularity score to the modularity score of its shuffled counterpart. Table 3 presents these modularity metrics for each sub-forum. When compared to the efficiency scores from Fig. 4(b), it is clear that the most modular sub-forums also tended to have the lowest diffusion efficiency (e.g. Readings and Study Groups) and vice versa (e.g. Questions for Professor and Course Material Feedback). Despite these trends, it is important to note the Technical Feedback forum, which had both low normalized modularity and diffusion efficiency (this sub-forum also had the lowest amount of participation and aimed to capture technical platform glitches instead of facilitating discussion about course content). Given the limited number of sub-forums over both MOOC instances, we refrain from reporting a Pearson r-value to quantify the anti-correlation between normalized modularity and diffusion efficiency. Successive iterations of this MOOC will offer additional opportunities and data to verify if the inverse relationships between modularity and efficiency scores observed in this investigation continue to hold and is an issue for further research.

Overall, these results reveal an important characteristic of discussion in MOOCs: when it comes to significant communication between learners, there is simply too many discussion topics and too pronounced heterogeneity characterizing participation to realize truly global-scale discussion. Instead, most information exchange, and by extension, any knowledge construction in the discussion forums occurs in small, short-lived groups.

## Discussion

Our research reveals that forum participation is heterogeneously distributed both temporally and according to the different discussion topics. User participation is driven by course milestones (for example, course launch and final project due date). Additionally, group dynamics – namely, the significance and vulnerability of communication – vary according to what is being discussed and how the forums are leveraged by course staff to encourage participation. Finally, modularity in MOOC forum networks appears to "trap" information in small learner groups. This finding is important as it highlights structural limitations that may adversely impact the ability of MOOCs to facilitate communication amongst learners that look to learn "in the crowd".

These insights into the communicative dynamics at play motivate a number of important questions about how social learning can be better supported, and facilitated, in MOOCs. Recent work on a subset of this data employed qualitative content analysis – combined with community detection schemes from machine learning – to infer latent learner communities according to the content of their forum posts. Interestingly, for the Cases and Final Projects sub-forums, the inferred communities had statistically significant differences in the geographic and prior educational experiences of constituent learners, as well as their final course performance and overall engagement in the discussion forums[25]. Moreover, ongoing semi-structured interviews of -FOBS-1 forum participants revealed different motivations for engaging in discussions – and different strategies related to managing "content overload" associated with the forums. These insights – coupled with those presented above – suggests the necessity of large-scale online learning platforms to increasingly leverage intelligent machine learning algorithms to support the needs of crowd-based learners. Such systems might, for example, detect different types of discussion and patterns of engagement during the runtime of a course to ultimately help students identify and engage in conversations that promote learning in accordance with individual objectives and learning preferences. Without such interventions – and the technical and pedagogical governance structures for online discussions that may result – the current structural limitations of social learning in MOOCs may prevent the realization of a truly global classroom.

# Methods

We analysed data from two successive instances of a course on business strategy (FOBS-1 and FOBS-2), offered on the Coursera platform in Spring and Autumn of 2013. Nearly 90,000 students registered for FOBS-1 and over 77,000 for FOBS-2. The course lasted for six weeks and assessed students through a combination of weekly quizzes and a final project: to perform a strategic analysis of any existing organization. Students were encouraged to use the forums to discuss weekly business cases on existing companies such as Google, Apple, Disney, etc. In FOBS-1, students were not evaluated on their performance in the forums; in FOBS-2, 8% of students' final scores was derived from their forum participation as a function of the total number of "upvotes" they received on their posts or comments.

In both instances, the discussion forum was segmented into sub-forums: the Final Project sub-forum facilitated questions, debates, and team formation for the final strategic analysis assignment; the Cases sub-forum was divided into additional sub-forums for each week's selected company, and each company's sub-forum was in turn divided into further sub-forums that asked a specific question about the company for that week (for example, questions about the companies competitive advantage); students posted for both logistical and content clarifications in the Questions for Professor sub-forum; Technical Feedback and Course Material Feedback sub-forums provided channels for voicing gratitude and felicitations at the end of the course, and suggestions for future improvements; Readings and Lectures harboured discussion around core course content; and the Study Groups sub-forum enabled learners to form cohorts with their peers to experience the course together – often interacting with others from similar timezones. Learners could create discussion threads within these sub-forums, which contained new posts or comments on existing posts.

We used social network analysis to capture both broad trends in communication and the roles of individuals in facilitating discussions. Network nodes represented learners that created at least one post or comment in a discussion thread; an edge connected two learners if they co-participated in at least one discussion thread. Formally, the communication network is represented as a graph $G = (V, E)$, where $V = \{s_i\}_{i=1}^{N}$ for the $N$ participants $s_i$, and $E = \{e_{ij}\}_{i,j=1; i \neq j}^{N}$ where $e_{ij} = m$ implies students $i$ and $j$ co-participated $m$ times in forum discussion threads. The networks presented here only depict explicit participation (e.g. posting or commenting), although the authors note that the prominence of lurking suggests the need for more robust network models that depict both viewing and posting trends.

**Deriving significant interaction networks**

The derivation included generating $K$ possible communication networks $\{G_i^f\}_{i=1}^{K}$ based on the trends in $N^f$ (i.e. the observed network for sub-forum $f$) and testing for the significance of each observed edge $e \in N^f$ with respect to its appearance in the sample graphs, using a one-sided z-test, $p < 0.001$ (here, the data's normality is justified by the central limit theorem due to the large number of edges). The Supplementary Information provides more details on the derivation of the significant networks and the criteria used to generate the sample graphs.

**Network diffusion**

To characterize the network diffusion efficiency, we simulated an SI (Susceptible/Infected) model[22] on the significant networks constructed from the sub-forums. The simulation first "infected" one node randomly, which then subsequently infected all neighboring nodes with the probability *p;* this process was repeated for all infected nodes. The outreach was measured as the ratio of the infected nodes to the total number of nodes over time. This number was averaged over a sample of 400 realizations with different initial seeds. The value of *p* was chosen to be 0.01; this choice did not impact the generality of the results (the same trends were obtained for different values of *p*).

## Tables

|  | FOBS -1 | | | FOBS -2 | | |
|---|---|---|---|---|---|---|
| **Sub-forum** | # parts. | # posts | posts/user (std. dev.) | # parts. | # posts | posts/user (std. dev.) |
| Technical Feedback | 250 | 519 | 2.08(2.8) | 109 | 233 | 2.14(3.3) |
| Course Material Feedback | 281 | 487 | 1.73(1.9) | 289 | 537 | 1,86(2.5) |
| Study Groups | 1,387 | 2,730 | 1.97(3.4) | 2,022 | 4,573 | 2.26(3.9) |
| Readings | 1,120 | 2,334 | 2.08(3.0) | 137 | 209 | 1.53(1.0) |
| Lectures | 638 | 1,262 | 1.98(2.8) | 470 | 1,175 | 2.50(3.0) |
| Questions for Professor | 305 | 610 | 2.00(3.1) | 370 | 785 | 2.12(3.2) |
| Final Project | 1,078 | 2,612 | 2.42(4.4) | 630 | 2,224 | 3.53(6.6) |
| Cases | 1,222 | 4,669 | 3.82(5.6) | 987 | 4,397 | 4.45(7.8) |

*Table 1: Summary statistics for forum participation in FOBS -1 and FOBS -2. # parts., #posts, and posts/user (std. dev.) denote the number of participants, number of posts, and average posts per user (along with standard deviations) for each sub-forum, respectively.*

|  | FOBS -1 | | FOBS -2 | |
|---|---|---|---|---|
| **Sub-forum** | # Orig. edges (# Nodes) | # Sig. edges(% decline) | # Orig. edges (# Nodes) | # Sig. edges(% decline) |
| Technical Feedback | 3,087 (231) | 339 (89%) | 767 (105) | 83 (89%) |
| Course Material Feedback | 2,752 (252) | 729 (74%) | 7,824 (279) | 1,253 (84%) |
| Study Groups | 41,819 (1,359) | 11,609 (72%) | 130,821(1,993) | 38,548 (71%) |
| Readings | 35,728(1,108) | 11,259(68%) | 1,873(137) | 125(93%) |
| Lectures | 12,644(617) | 3,988(68%) | 6,839(463) | 3,472(49%) |
| Questions for Professor | 2,758(284) | 896(68%) | 3,258(345) | 1,358(58%) |
| Final Project | 23,244(1,019) | 12,557(46%) | 16,116(611) | 7,760(52%) |
| Cases | 102,171(1,114) | 57,490(44%) | 45,257(925) | 27,586(39%) |

Table 2: Observed and derived communication networks for the different sub-forums in FOBS -1 and FOBS -2.

| | FOBS -1 | | | FOBS -2 | | |
|---|---|---|---|---|---|---|
| **Sub-forum** | Original | Shuffled | Normalized Modularity | Original | Shuffled | Normalized Modularity |
| Technical Feedback | 0.78 | 0.55 | 1.42 | 0.86 | 0.72 | 1.19 |
| Course Material Feedback | 0.66 | 0.36 | 1.83 | 0.47 | 0.26 | 1.81 |
| Study Groups | 0.80 | 0.19 | 4.21 | 0.71 | 0.12 | 5.92 |
| Readings | 0.70 | 0.17 | 4.11 | 0.43 | 0.15 | 2.87 |
| Lectures | 0.61 | 0.22 | 2.77 | 0.51 | 0.22 | 2.32 |
| Questions for Professor | 0.65 | 0.36 | 1.81 | 0.62 | 0.32 | 1.94 |
| Final Project | 0.58 | 0.15 | 3.87 | 0.42 | 0.15 | 2.80 |
| Cases | 0.27 | 0.10 | 2.70 | 0.34 | 0.10 | 3.40 |

Table 3, Modularity scores of the original and shuffled networks.

# Figures

Figure 1 – The observed (a) and derived (b) communication networks for the Study Groups sub-forum. Here, we can see the impact of link filtration on network properties such as modularity score, which equals 0.62 and 0.80 for a and b, respectively. Colours correspond to the detected communities.

Figure 2 – Network vulnerability for the different sub-forums in FOBS-1. Here, "LCC" refers to the largest connected component in each sub-forum's communication network.

Figure 3 – Vulnerability versus random removal for Cases sub-forum in FOBS-1. Here, "LCC" refers to the largest connected component in this sub-forum's communication network.

Figure 4 (a) shows the percentage of infected nodes vs. simulation time for different networks. The solid lines show the results over the original network and the dashed lines for the degree-preserved shuffled network (configuration model), and (b) shows the value of e for different sub-forums.

# Acknowledgements

The authors would like to thank the MOOC Research Initiative for its funding and support. Many thanks to Michael Osborne and Steve Roberts in the University of Oxford's Department of Engineering for their insights, as well as Coursera and the University of Virginia for their assistance with acquiring and navigating the dataset.

## Author Contributions
RE, IH, NG, and TY contributed equally to writing and reviewing the text of the manuscript. NG and TY implemented the computational models presented in the work and generated all relevant figures and tables.

## Additional Information

### Competing financial interests
The authors declare no competing financial interests.

### Supplementary Information
1. Supplementary pdf
2. Supplementary video

# Figures

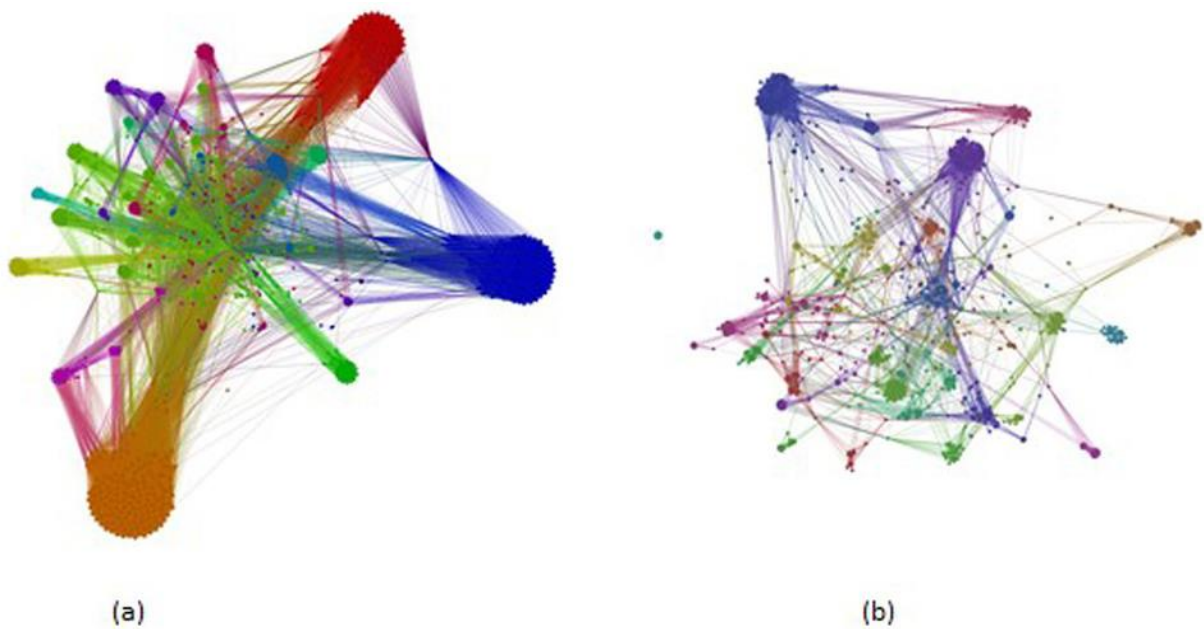

*Figure 1 – The observed (a) and derived (b) communication networks for the Study Groups sub-forum. Here, we can see the impact of link filtration on network properties such as modularity score, which equals 0.62 and 0.80 for a and b, respectively. Colours correspond to the detected communities.*

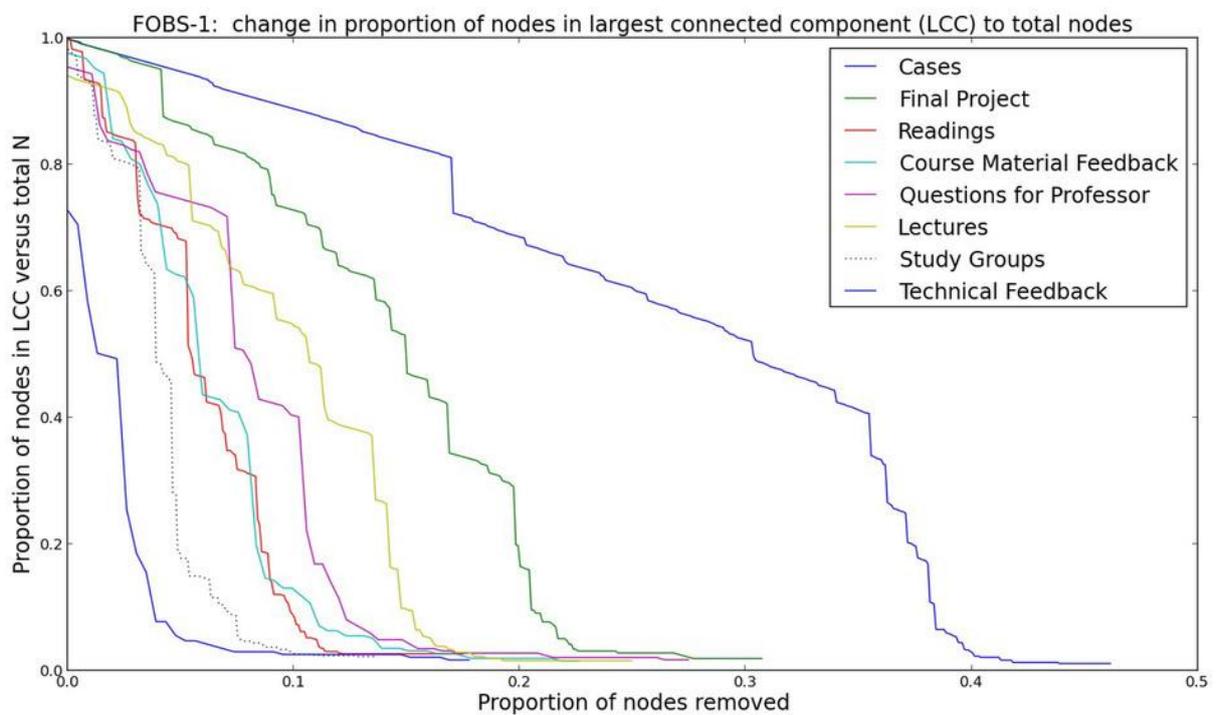

*Figure 2 – Network vulnerability for the different sub-forums in FOBS-1. Here, "LCC" refers to the largest connected component in each sub-forum's communication network.*

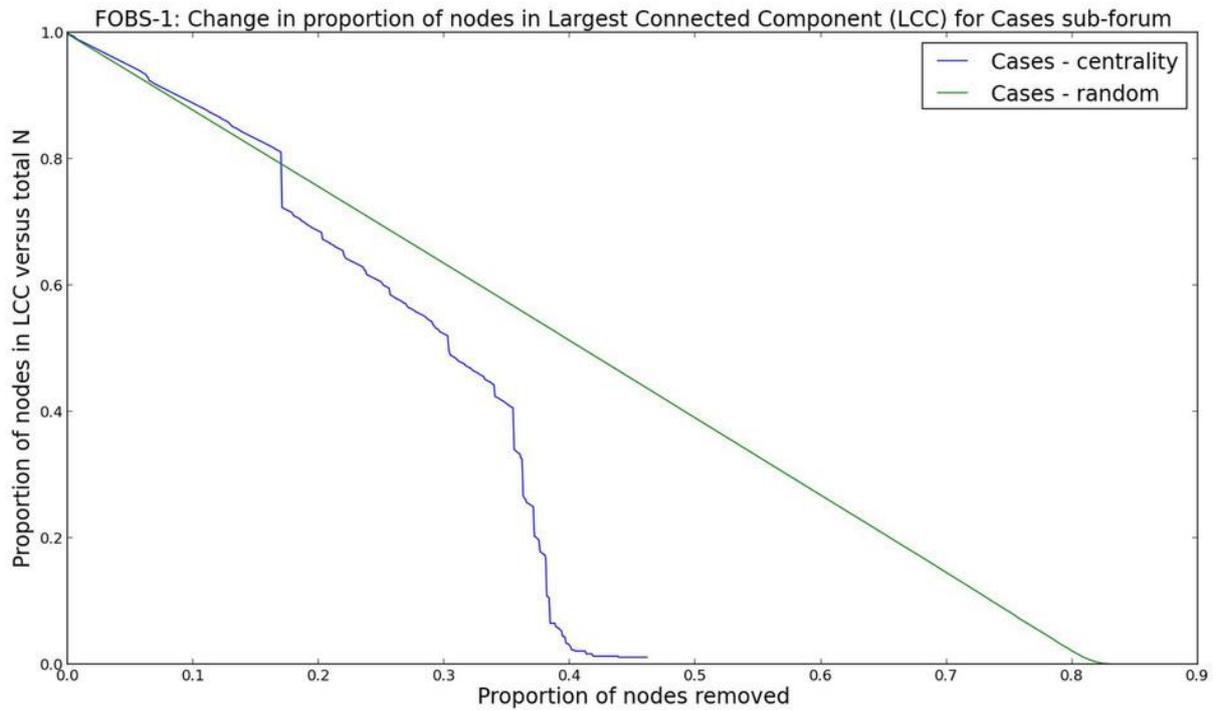

*Figure 3 – Vulnerability versus random removal for Cases sub-forum in FOBS-1. Here, "LCC" refers to the largest connected component in this sub-forum's communication network.*

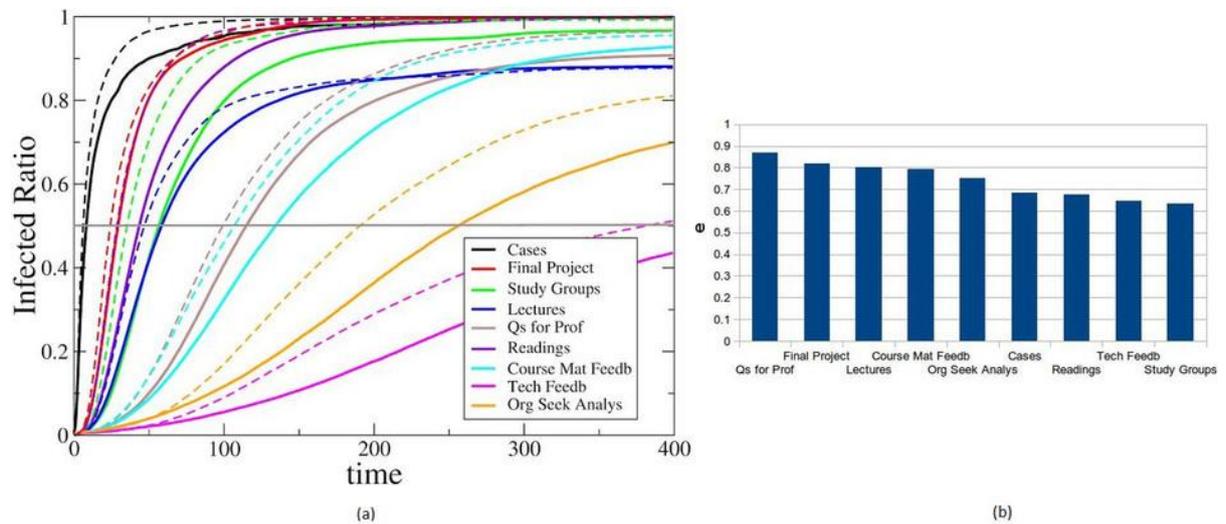

*Figure 4 (a) shows the percentage of infected nodes vs. simulation time for different networks. The solid lines show the results over the original network and the dashed lines for the degree-preserved shuffled network (configuration model), and (b) shows the value of e for different sub-forums.*

# Supplementary Information for: *Structural limitations of learning in a crowd: communication vulnerability and information diffusion in MOOCs*

Nabeel Gillani [1], Taha Yasseri [2], Rebecca Eynon* [2], Isis Hjorth [2]

[1]
Department of Engineering Science
University of Oxford
Parks Road
Oxford OX1 3PJ
UK
nabeel@robots.ox.ac.uk

[2]
Oxford Internet Institute
University of Oxford
1 St. Giles'
Oxford OX1 3JS
UK
{taha.yasseri, rebecca.eynon, isis.hjorth}@oii.ox.ac.uk

**Derivation of significant social network**

Defining how nodes should be connected - i.e. the role of the edge set - was not immediately obvious. Previous work has employed a number of different topological definitions – both directed and undirected - for modelling discussions in online forums[1]. Because students sometimes tended to create new posts even when they meant to comment, we determined that a directed topology would not adequately capture the potential information flow and communication between learners.

Our guiding philosophy in formulating the network was to make the least assumptions about our data to determine which links depicting thread co-participation were "significant" (i.e., which ones were indicative of exchanges between learners that could indicate the existence, or potential for future existence, of underlying social relationships). Past research has used interaction time windows to determine which links to keep and which to remove in a network[2]. Because we did not have reliable a priori knowledge about what a reasonable time window would be, we instead turned to a significant network extraction model used to infer social networks in ecological settings[3] in order to inform our efforts.

The derivation of a significant social network proceeds as follows. For each sub-forum $f$, we first construct a bipartite graph mapping and represented by the *learner-to-thread* adjaceny matrix $B_f^{N_f \times T_f}$, where $N_f$ represents the number of learners that explicitly posted and $T_f$ the number of threads, both in sub-forum $f$. Each entry $b^{n,t} \in B_f^{N_f \times T_f}$ is an integer greater than or equal to 0, denoting the number of times learner $n$ participated in thread $t$. Next, we compute a standard weighted one-mode projection of $B_f$ to recover the *learner-to-learner* network, $L_f^{N_f \times N_f}$ [3]. Each entry $l^{i,j} \in L_f^{N_f \times N_f}$ is also an integer greater than or equal to 0, depicting the sum over minimum number of times that learner $i$ or $j$ participated in a particular thread $t$, i.e. $l^{i,j} = \sum_{t \in T_f} min(b^{i,t}, b^{j,t})$.

With a learner-to-learner network $L_f$, we are now tasked with identifying which edges in $L_f$ (i.e., $l^{i,j} \in L_f$ s.t. $l^{i,j} \neq 0$) depict a *significant* interaction between two learners. Our goal is to generate a family of $M$ sample networks against which we can test the significance of each $l_{i,j} \in L_f$. We start by noting that the observed learner-to-thread network $B_f$ depicts each learner $n$'s participation in a particular thread $t$. We can model this participation – i.e., each row $n$ of $B_f$ – as a draw from $Multinomial(k_n, p_n)$, where $k_n = \sum_{t \in T_f} b_{n,t}$ and $p_n = (p_{n,1}, \ldots, p_{n,T_f})$ for $p_{n,t} = b_{n,t}/k_n$. It is important to note that $p = \{p_n\}_{i=1}^N$ represents the observed social relationships between learners as represented by the likelihood of each student's participation in a particular thread.

If we wish to test the significance of the observed edges – i.e., the observed social interactions – we must determine a mechanism for generating possible social networks that do not possess the same social patterns as the observed one. In order to explore alternative social structures, we first define a shuffling function $\sigma$ such that each row of the $s$th sample learner-to-thread network $B_f^s$ is drawn from $Multinomial(k_n, \sigma(p_n))$ with $k_n$ and $p_n$ defined as above. We define $\sigma$ such that it preserves learner $n$'s proportional participation in different threads (e.g., the entropy of each $p_n$), but accounts for the possibility of participation in alternate threads. As an extension to Psorakis et al., we constrain $\sigma$ to only shuffle each entry of $p_n$ with a location (e.g. thread) $t$ that has popularity greater than or equal to the least popular thread that learner $n$ participated in, where thread popularity is defined as the number of posts it contains. This constraint is meant to reflect which threads learners could have possibly participated in, since in many cases, discussion threads only had a single or very small number of posts, and therefore, it is unrealistic to assume that a learner who participated primarily in popular threads may have also participated in isolated ones. Without this constraint, the shuffling allocates participation probabilities to a larger set of threads, increasing the likelihood – particularly for those individuals with low participation volumes but high proclivity to post in popular threads – that these one-off interactions are deemed "significant". Additionally, this constraint is informed by real-world discussions with participants from FOBS-1, some of who indicated that the popularity of a particular discussion thread often influenced their decisions to view or post. Therefore, the constrained shuffling more accurately captures learner behaviour and detects one-off participation in high-activity threads as insignificant (thereby, pruning more edges) when compared to its under-constrained counterpart.

With a sampling procedure in place, we generate each $B_f^s$ and compute its one mode projection to arrive at the set of sampled learner-to-learner networks, i.e. $G = \{G_f^i\}_{s=1}^M$. We can then compare each entry $l_{i,j} \in L_f$ to $\frac{1}{M}\sum_s g_{i,j}$ for $g_{i,j} \in G_f^s$, computing the z-score and labeling as significant if the right-tailed p-value is less than 0.001 (we assumed a relatively small p-value threshold due to the sparsity of participation in the discussion threads). Our derived significant network is the collection of $l_{i,j} \in L_f$ labeled as significant by this procedure.

It is important to note that the determination of social significance by this procedure relies almost entirely on the frequency with which learners co-participate in threads. Frequency of forum participation has been investigated by others as a means of determining engagement and evaluating performance in educational settings[2]. Still, in a complex social setting such as an online course, the significance of communication is not entirely dependent on the frequency of co-participation, but also on the nature of the content exchanged. Developing automated ways of content-based significance testing is an opportunity for further research in MOOCs.

**Supplementary figures**

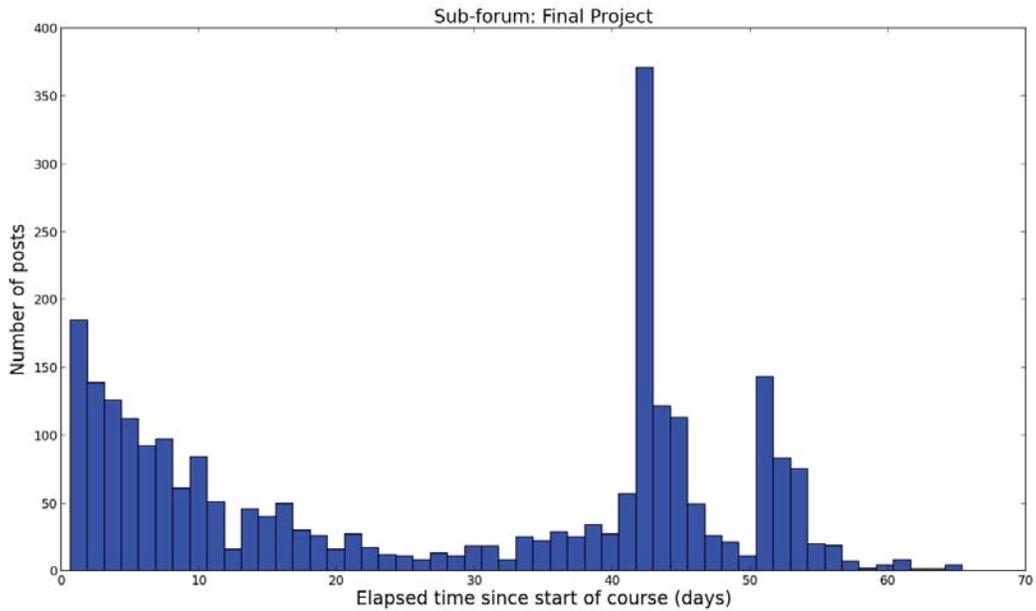

Figure S1: Forum post activity over time in the Final Project sub-forum of FOBS-1. The large peak around the end of week 6 corresponds to students posting last-minute questions about the final project submission deadline.

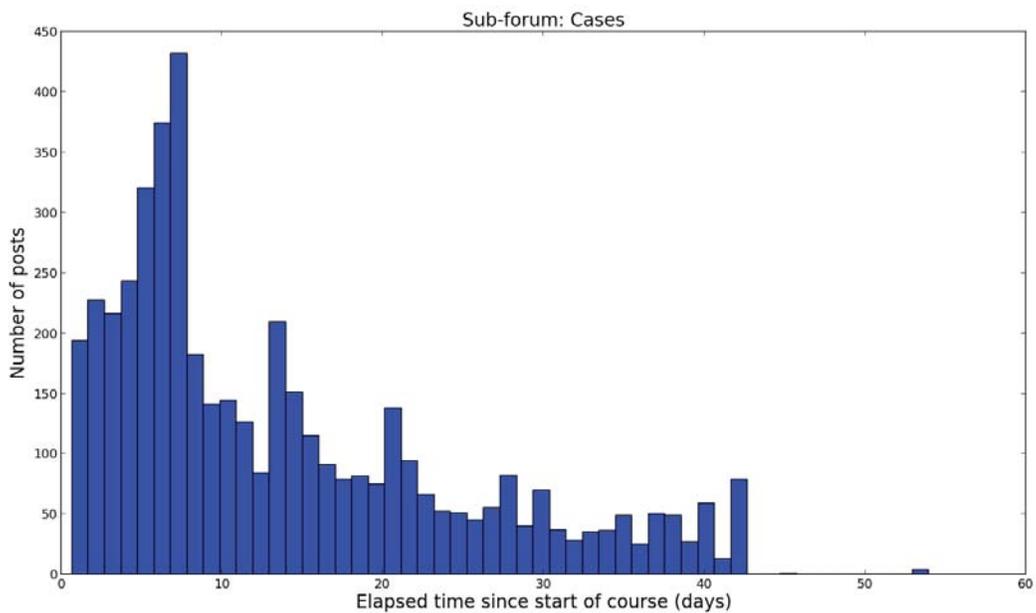

Figure S2: Forum post activity over time in the Cases sub-forum of FOBS-1. Like many of the other sub-forums, participation decreases as the course progresses, but there are still peaks of activity each week corresponding to the weekly case discussions.

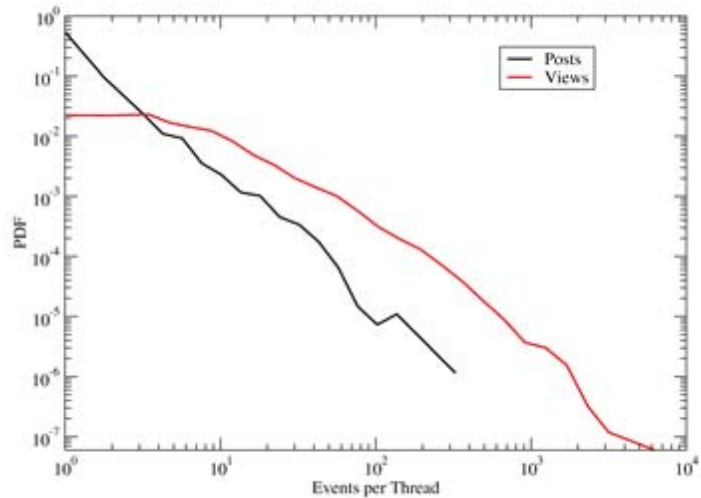

Figure S3: The number of views and posts per discussion threads across all sub-forums, in log-log scale, for FOBS-1. The charts suggest a fat-tailed distribution of views and posts across threads – i.e., the vast majority of discussion threads have very small numbers of posts and views, with a few threads harbouring high posting and viewing behaviour.

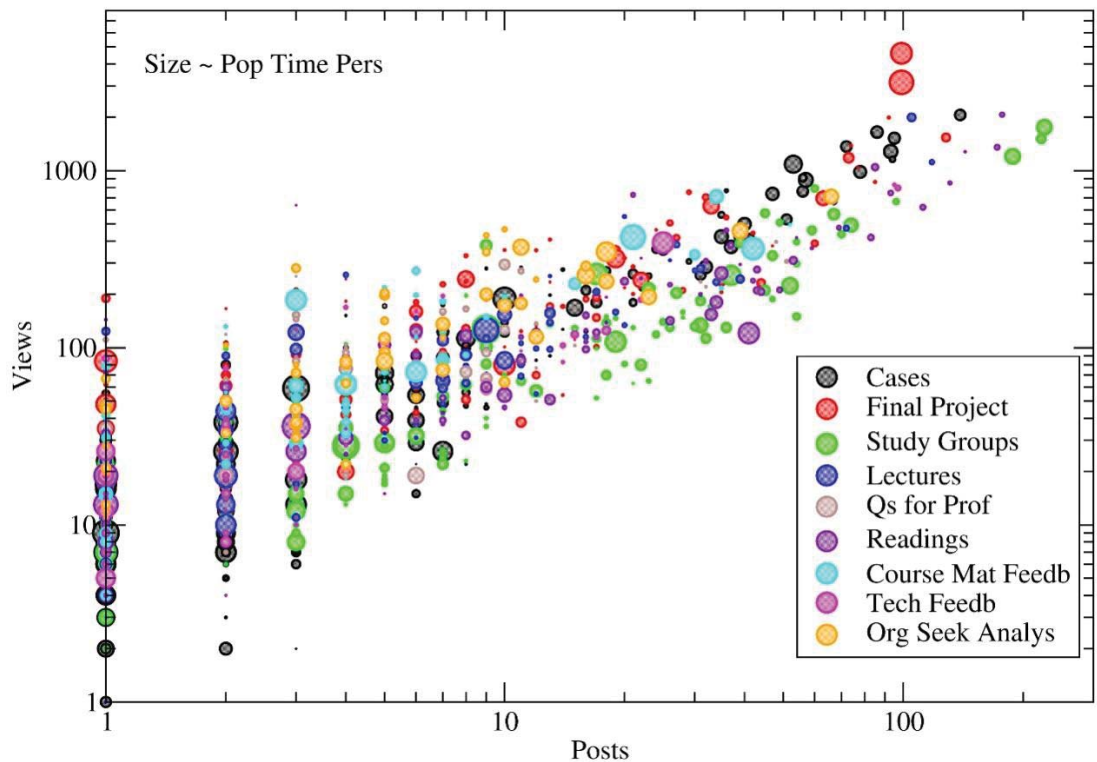

Figure S4: Comparison of posts and views for each thread in a particular sub-forum, denoted by the coloured circles shown here, for FOBS-1. The size of each circle indicates the "Popularity time persistence" of the corresponding thread, i.e., the amount of time that elapses before 90% of all posts are made to that thread (hence, small circles depict threads with very short lifespans).

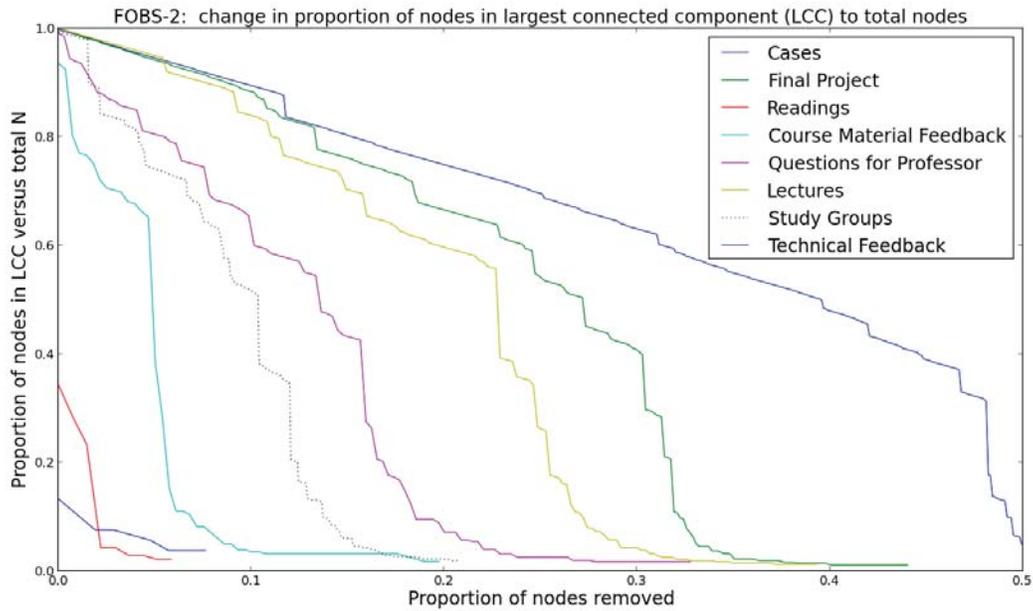

Figure S5: Communication vulnerability in the different sub-forums of FOBS-2. These trends are similar to those observed in FOBS-1.

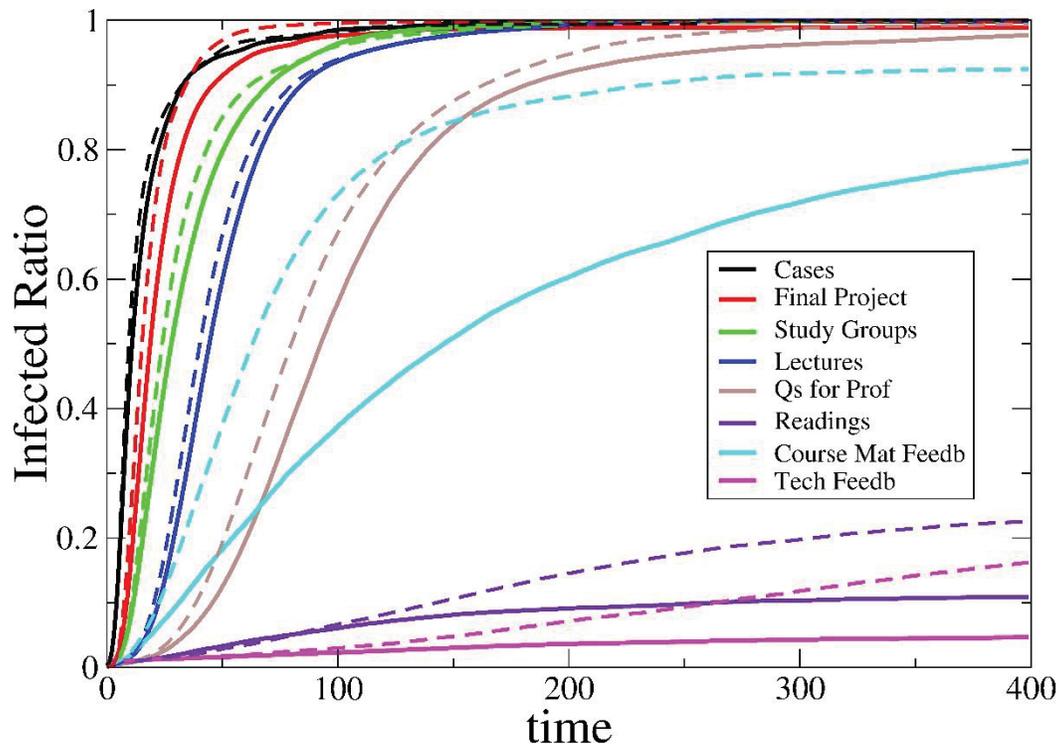

Figure S6: shows the percentage of infected nodes vs. simulation time for different networks in FOBS-2 (similar to those observed in FOBS-1). The solid lines show the results over the original network and the dashed lines for the degree-preserved shuffled network (configuration model).

**Supplementary video**

The video at the link below depicts the network vulnerability simulation for the Final Project sub-forum from FOBS-1. Each frame corresponds to one step in the algorithm, at which the node with the highest betweenness centrality is computed and disconnected from the graph.

https://www.dropbox.com/s/rvkd18dnuiyd02v/finalprojects.avi